# Interaction between DNA and the Hfq amyloid-like region triggers a viscoelastic response

Omar El Hamoui[a,†,‡], Indresh Yadav[b,‡], Milad Radiom[c], Frank Wien[a], Jean-Francois Berret[c], Johan R. C. van der Maarel[b] and Véronique Arluison[d,e*]

[a] *Synchrotron SOLEIL, F-91192 Gif-sur-Yvette, France*
[b] *Department of Physics, National University of Singapore, Singapore 117542, Singapore*
[c] *Matière et Systèmes Complexes, UMR 7057 CNRS Université de Paris, Bâtiment Condorcet, 10 rue Alice Domon et Léonie Duquet, F-75205 Paris, France*
[d] *Laboratoire Léon Brillouin LLB, CEA, CNRS UMR12, Université Paris Saclay, CEA Saclay, F-91191 Gif-sur-Yvette, France*
[e] *Université de Paris, F-75006 Paris, France*

‡ *Equivalent contribution*
** Corresponding author: Véronique Arluison; LLB UMR 12 CEA/CNRS, CEA Saclay, 91191 Gif-sur-Yvette Cedex, France; Tel 33 (0)1 69 08 32 82; Email: veronique.arluison@u-paris.fr*
† *Present Address: Omar El Hamoui, Institut des Sciences Analytiques et de Physico-chimie pour l'Environnement et les Matériaux, UMR 5254 CNRS/Université de Pau et des Pays de l'Adour, 2 Avenue du Président Pierre Angot, 64000 Pau, France*

**ABSTRACT**

Molecular transport of biomolecules plays a pivotal role in the machinery of life. Yet, this role is poorly understood due the lack of quantitative information. Here, the role and properties of the C-terminal region of *Escherichia coli* Hfq is reported, involved in controlling the flow of a DNA solution. A combination of experimental methodologies has been used to probe the interaction of Hfq with DNA and to measure the rheological properties of the complex. A physical gel with a temperature reversible elasticity modulus is formed due to formation of non-covalent crosslinks. The mechanical response of the complexes shows that they are inhomogeneous soft solids. Our experiments indicate that Hfq C-terminal region could contribute to genome's mechanical response. The reported viscoelasticity of the DNA-protein complex might have implications for cellular processes involving molecular transport of DNA or segments thereof.

# INTRODUCTION

Cells are highly complex superstructures, made of different subcellular structures surrounded by one or two membranes (*i.e.* for Gram-negative bacteria). Eukaryotic cells for instance contain organelles, such as mitochondria, Golgi apparatus and nucleus. The nucleus is the largest or-ganelle that encloses the genetic information. Nucleus content is separated from the cytoplasm by the nuclear membrane, inside which the DNA is shaped by proteins to form the compacted chromatin. DNA is transcribed into RNA in the nucleus, exported to the cytoplasm through nuclear pores, and eventually translated into proteins. Bacteria lack a membrane-surrounded nucleus, and DNA, RNA and proteins are not separated in the cytoplasm. The advantage of this arrangement is that it has a direct consequence on the speed of the transfer of genetic information. Moreover, the processes of replication, transcription, and translation all occur within the same cellular compartment. However, despite the absence of intracellular membranes, bacteria have an intracellular spatial-organization[1]. For instance, they have a bacterial cytoskeleton responsible for cell division[2] or nanostructures dedicated to RNA metabolism[3]. One of the most complex structures of the bacterial cell is probably its genome, a circular chromosomic DNA. The bacterial genome is highly compacted and forms a structure called nucleoid[4]. In terms of dimensions, it is estimated that the 1 mm-long bacterial genomic DNA is compacted to a nucleoid volume of ~ 0.15 $\mu m^3$[3, 5]. Random coil chromosomal DNA would occupy a volume of ~ 500 $\mu m^3$[3, 4]. Several factors affect DNA compaction and the formation of the nucleoid, including: *(i)* molecular crowding involving DNA segregation through overall repulsion between components of the system[6, 7]; *(ii)* supercoiling of genomic DNA (in *E. coli*, DNA is negatively supercoiled)[6, 8]; *(iii)* RNA stabilization[4, 9]; *(iv)* short and long-range interactions mediated by proteins[10]. The action of different proteins, including Nucleoid Associated Proteins (NAPs) strongly influence DNA compaction[10]. In *E. coli*, NAPs are abundant and represent 50000 to 200000 proteins per cell[11]. They are known to bend or bridge DNA and/or to form filaments on DNA[10, 12, 13]. The resulting effect of NAPs is that the nucleoid is composed of topologically dynamic domains forming toroids, filaments or plectonemes[14]. Thus, the nucleoid is far from being randomly organized; it is fairly structured and includes linear ordering and local alignments[15, 16]. This structuring induces DNA mechanical tensions that are connected to genetic regulation and affects certain regulatory processes[17-20].

Indeed, to function properly, sigma factors and polymerases, among other proteins, must diffuse through this densely packed DNA to find their binding sites. For this reason, too much supercoiling and constraints are harmful for the cell as they may block transcription, replication or chromosome segregation. Proteins involved in these processes find specific sites on DNA by a combination of 1D-diffusion along the DNA molecule and 3D diffusion through the immersing medium from a particular binding site to another site (hopping and jumping)[18, 21]. The resulting diffusion constant of a protein in a bacterium such as *E. coli* is usually about 0.4 $\mu m^2/s$ with the highest mobility in the nucleoid region[22, 23]. Protein diffusion depends on crowding, degree of confinement, and spatial organization of the genome[6, 7]. Here, we show for the first time that a part of a bridging-protein (*i.e.* the C-terminal region of Hfq) directly changes DNA's viscoelasticity. We surmise that this result provides a new role for NAPs on target search, because protein diffusion involves hydrodynamic interaction and flow of the genome by, among others, moving DNA segments out of the way[4].

Our analysis focuses on an NAP called Hfq, found in about half of bacterial species[24]. Since its identification in RNA bacteriophage Qβ replication (which gives its name Host Factor for bacte-riophage Qβ[25]), many functions have been identified for Hfq, mainly related to RNA metabolism[26]. Indeed, Hfq exhibits a pleiotropic role, covering a wide range of functions in bacterial survival[27]. These include resistance to various stresses, influence on growth rate, pathogenesis, mutagenesis rate and in plasmid compaction and supercoiling[26-31]. Most of these phenotypes could be explained by the involvement of Hfq in small non-coding RNA (sRNA) based regulation. Hfq promotes the pairing of sRNAs to their target mRNAs, mediating post-transcriptional regulation (mRNA stability and translation efficiency) [32, 33]. But in addition, Hfq interacts also with DNA and plays a role in bacterial DNA compaction, although its distribution within the nucleoid might be more heterogeneous than the most common NAPs H-NS and HU[31, 34]. Therefore, while the overall minimal concentration of nucleoid Hfq is about 10-15 µM[35, 36], its concentration must vary locally from tens to hundreds of µM, with important consequences on DNA compaction and alignment[16, 30]. Some of the phenotypes of *hfq* mutant cells, such as change in growth rate and mutation sensitivity, may be related to its DNA binding properties.

Structurally, Hfq is related to the Sm protein family that are also involved in RNA-related processes such as splicing and RNA degradation[37]. Hfq forms a toroidal ring similar to Sm proteins, except that in bacteria the torus consists in homohexameric oligomers, whereas in archeaes and eukaryotes it tends to be formed by homo or heteroheptameric oligomers. The amino-terminal region of Hfq (NTR, about 65 aminoacids residues) is formed by an antiparallel β-sheet followed by an N-terminal α-helix; the NTR bent β-sheets of each monomer interact in order to form the hexameric torus. Both faces (called proximal and distal faces) and the rim of the toroidal hexamer can contact nucleic acid, but DNA is rather bound on the proximal face, *i.e.* the surface with the N-terminal α-helices[38, 39]. On the other hand, the C-terminal region (CTR) of Hfq comprises about 40 amino acid residues outside of the Sm NTR ring, although no 3D structure is known for this CTR. Nevertheless, the CTR also binds DNA and it has been shown that it self-assembles into an amyloid-like structure *in vitro* and *in vivo*[40, 41]. Recently, this CTR region has been shown to play a major role in DNA bridging and compaction[30, 31], but does not affect directly the overall DNA topology[30]. In this work, we investigate a new property of the amyloid-like region of Hfq, specifically its ability to change the properties of the flow of DNA. We have opted to investigate the CTR rather than full length Hfq, thereby avoiding complications related to changes in secondary structure and partial melting of double stranded DNA[42, 43]. Plausible functional consequences on gene expression will also be discussed.

# EXPERIMENTAL SECTION

## *Hfq C-terminal (CTR) peptide synthesis and preparation*

The peptide corresponding to the amyloid CTR domain of Hfq (residues 64 to 102, referred to as Hfq-CTR throughout the manuscript) was synthetized by Proteogenix SA (France). The sequence of the Hfq-CTR peptide is SRPVSHHSNNAGGGTSSNYHHGSSAQNTSAQQDSEETE. Hfq-CTR was reconstituted in water at 20 mg/mL. The self-assembly into large fibers is not instantaneous, and the time needed to observe aggregation is dependent on the temperature and the presence of DNA[30] (in full length Hfq, the CTR of individual subunits are attached to the tore and such a high concentration or time may not be necessary). As a negative control, a mutant peptide unable to bind DNA and consequently self-assemble (see below) within the experimental timeframe was also analyzed. Its sequence is SRPVSHHSNNAGGGTAANYHHGSSAQNTSAQQDSEETE (Hfq-CTRS80,81A).

### *Preparation of the complexes for SRCD and microrheology*

Complexes between Hfq-CTR peptides and $(dA:dT)_{59}$ DNA duplex (Eurogentec) were prepared as previously described[16, 30]. We chose this $(dA:dT)_{59}$ DNA sequence because CTR has the best affinity for AT-rich sequences (~ 250 nM[43]). It is important to avoid using longer DNAs as they may partially align naturally without Hfq-CTR[16, 44]. In our experiments, the DNA:CTR stoichiometry varies and is always expressed in Hfq-CTR peptide and base pair (bp) concentrations (Table I). Samples were analyzed at specific times, usually after about 2 weeks to allow peptide self-assembly on DNA that might not be instantaneous[30].

| DNA $(dA:dT)_{59}$ (mM in bp) | CTR (mM) | DNA:CTR stoichiometry | Technique | Apparent rheological Behavior |
|---|---|---|---|---|
| 7.3 | 0 | 1:0 | wire, bead | viscous liquid |
| 1.8 | 1.8 | 1:1 | bead | Gel |
| 7.8 | 3.9 | 2:1 | bead | Gel |
| 7.3 | 1.8 | 4:1 | wire, bead, SRCD | Gel |
| 0 | 1.8 | 0:1 | wire | viscous liquid |

***Table I:*** *List of DNA/CTR samples studied and experimental conditions used.*

### *Wire based microrheology*

The magnetic wire microrheology (µrheology) technique has been described previously[45-47]. In brief, wires were synthesized by electrostatic co-assembly of iron oxide nanoparticles (NPs) and with poly(diallyldimethylammonium chloride) (PDADMAC, Aldrich, $M_w$ > 100000 g mol$^{-1}$)[48]. NPs of size 6.7 nm and 13.2 nm were used for the synthesis, resulting in wires with different magnetization properties. Low viscosity fluids such as DNA and CTR solutions were studied with wires made with the smaller NPs, whereas experiments of the DNA-CTR gels were conducted with wires made from the larger ones. The geometrical characterization of the wires was performed by measuring the length $(L)$ and the diameter $(D)$ of >100 wires using a 100× objective on an optical microscope (Olympus IX73) coupled with a CCD camera (QImaging, EXi Blue) supported by the software Metaview (Universal Imaging). From the distribution, we determined the reduced wire length, $L^* = L/\left[D\sqrt{g(L/D)}\right]$, where $g(x) = ln(x) - 0.662 + 0.917/x - 0.050/x^2$[45, 49]. The magnetic wires used in this study have lengths between 10 and 100 µm and diameters between 0.6 and 2 µm. We observed that the wire diameter slightly increases with $L$, probably due to bundle formation. As a result, scaling laws of the form $L^*(L) = \alpha L^{\beta}$ were found with $(\alpha, \beta)$ = 1.92 and 0.615 for the 6.7 nm NPs and $(\alpha, \beta)$ = 1.66 and 0.638 for the 13.2 nm NPs. A volume of 0.5 µL containing $10^5$ wires in phosphate buffer saline PBS (pH 7.4) was then added to the DNA, CTR or DNA-CTR dispersions and gently stirred. A volume equal to 25 µL of this disper-sion was then deposited on a glass plate and sealed into a Gene Frame® (Abgene/Advanced Biotech, dimensions 10×10×0.25 mm$^3$). The glass plate was introduced into a homemade device generating a rotational magnetic field, from two pairs of coils (resistance 23 Ω) working with a 90°-phase shift. An electronic set-up allowed measurements in the frequency range $\omega$ = 10$^{-2}$ – 10$^2$ rad s$^{-1}$ and at a magnetic field $\mu_0 H$ = 10.3 mT. The µrheology protocol used is based on the Magnetic Rotational Spectroscopy technique[46, 50, 51]. For each condition of magnetic field and angular frequency, a movie was recorded for a period of time of at least 10/$\omega$ and then treated using the ImageJ software. For calibration, experiments were performed in water-glycerol solutions of viscosities between 5 and 100 mPa s at room temperature, leading to a susceptibility anisotropy coefficient $\Delta\chi$ = 0.054 ± 0.006 for the 6.7 nm NPs and $\Delta\chi$ = 2.3 ± 0.5 for the 13.2 nm NPs[52, 53].

***Passive microrheology (particle video tracking)***

Particle tracking experiments were carried out with a Nikon Eclipse Ti-U microscope equipped with a 100x long working distance objective. The temperature was controlled with a Linkam heating stage. The samples were spiked with polystyrene microspheres (Polysciences, Warrington, PA) of 1.93 ± 0.05 µm diameter with a final concentration of 0.005 wt %. A sample having a volume of 10 µL was deposited on a microscopy slide and sealed with a cover slip separated by a spacer of thickness 0.12 mm and cavity diameter 1 cm. In order to minimize hydrodynamic interaction, the selected beads for imaging were separated by at least 10 bead diameters (20 µm). Furthermore, the height level of the focal plane was adjusted so that it is situated right between the slide and cover slip with maximum separation. The trajectories of 10 -25 different beads in randomly chosen locations were recorded with a metal oxide semiconductor (CMOS) camera (Basler A504k) at a rate of 250 frames per second. Video clips of 5 min duration were analyzed with MATLAB (Natick, MA) and the particle trajectories were obtained with public domain tracking software (http://site.physics.georgetown.edu/matlab/). All further data analysis was done with home-developed software scripts written in MATLAB code. The pixel resolution of 0.12 µm was calibrated with the help of a metric ruler. We have checked our setup by monitoring immobilized beads adsorbed to a glass slide. These beads exhibited a mean-square displacement of 10$^{-5}$ µm$^2$, which sets a limit to the lowest measurable creep compliance $J$ of 0.02 m$^2$/N.

***Electrophoretic Mobility Shift Assay (EMSA)***

The binding of Hfq–CTRs to DNA was investigated with EMSA. The DNA fragment was incubated with Hfq–CTR (WT and mutant) at room temperature for 20 min. EMSAs were carried out using non-denaturing gradient 4–12% polyacrylamide gel (Bio-rad). The native gel was run for 2 h at room temperature with 40 mM Tris-Acetate, 1 mM ethylenediaminetetraacetic acid, pH 8.0 (TAE) buffer, stained with GelRed nucleic acid stain (Biotium), and imaged with a G-BOX system (Syngene, Cambridge, UK).

***Synchrotron Radiation Circular Dichroism (SRDC)***

For SRCD analysis, measurements and data collection were carried out on DISCO beam-line at the SOLEIL Synchrotron (proposal 20171061)[54]. 2-4 µl of samples were loaded into circular demountable $CaF_2$ cells of 33 µm pathlength[55]. Three separate data collections with fresh sample preparations were carried out to ensure consistency and repeatability. Spectral acquisitions of 1 nm steps at 1.2 integration time, between 320 and 180 nm were performed in triplicate for the samples as well as for the baselines. (+)-camphor-10-sulfonic acid (CSA) was used to calibrate amplitudes and wavelength positions of the SRCD experiment. Data analyses including averaging, baseline subtraction, smoothing, scaling and standardization were carried out with CDtool[56].

# RESULTS AND DISCUSSION

## Binding of Hfq-CTR on DNA

We first confirmed the interaction of the Hfq-CTR peptide and our model $(dA:dT)_{59}$ DNA with Electrophoretic Mobility Shit Assay (EMSA) and Synchrotron Radiation Circular Dichroism (SRCD). The EMSA result is shown Figure 1A. A clear band shift is observed in the case of the Wild Type (WT) Hfq-CTR, confirming strong binding[16, 30]. No band shift is observed upon addition of the Hfq-CTRS80,81A mutant. This indicates that the CTR mutant does not bind DNA, at least to a degree which can be detected with the EMSA assay. Binding by WT Hfq-CTR and absence of binding by Hfq-CTRS80,81A mutant to DNA were confirmed with SRCD. The spectra pertaining to WT-CTR and mutant S80,81A-CTR are shown in Figure 1B and Figure 1C, respectively. Spectra of the individual components as measured from solutions with equivalent DNA and CTR concentrations are also displayed. For WT-CTR, significant spectral changes in the 220 nm region (inversion of CD signal) and 260-280nm spectral band are observed. These correlate with previously identified amyloid formation, base-tilting and base-pairing of AT rich sequences, respectively[16]. Furthermore, the relevant combination of the spectra pertaining to the individual components does not match the spectrum of the WT-CTR/DNA mixture, the theoretical summation of the individual spectra does not result in the experimentally measured spectra of the complexed CTR-DNA. The observed spectral changes for WT-CTR have been associated with amyloid formation[16, 30]. In contrast, following addition of S80,81A-CTR to DNA no significant differences in the SRCD spectrum were observed (Figure 1C). Here, the experimentally observed spectrum is close to the recomposed spectrum obtained from the relevant combination of the individual components. Any small difference indicates that the interaction between S80,81A-CTR and DNA must be non-existent or very weak. Hydrogen-bonding of S80,81A-CTR on DNA is strongly reduced and amyloid-like polymerization is suppressed by the change from Serine to Alanine amino acid residues[30].

Note that in the cell, the CTRs are attached to the toroidal assembly formed by the full-length protein. Thus they may behave differently [57]. Nevertheless full-length Hfq and CTR-Hfq displayed similar SRCD signals (specifically an inversion between 220-240 nm indicative for base tilting, not shown), suggesting that full length-Hfq and Hfq-CTR behave similarly. In contrast, for the Hfq-CTRS80,81A mutant-DNA complex no such spectral differences could be observed. The experimental spectrum of the mutant complexed to DNA resembled the theoretical sum of the isolated mutant Hfq-CTR and DNA spectra. The subtle difference indicates that the interaction must be nonexistent (or rather weak). This is certainly due to the fact that within in the DNA binding region of Hfq-CTR the change of amino acid side chains from Serine to Alanine does not allow the formation of hydrogen-bonding involved in the binding of the CTR to DNA, which strongly reduces the affinity[58]. These results can be explained by the fact that the Hfq-CTR mutant is not changing its structure, as the DNA cannot promote the polymerization (amyloid-like) of the Hfq-CTR[30]. Indeed, the spectral signature at 220 nm of amyloid formation coinciding with the DNA tilting is lacking for the mutant Hfq-CTR.

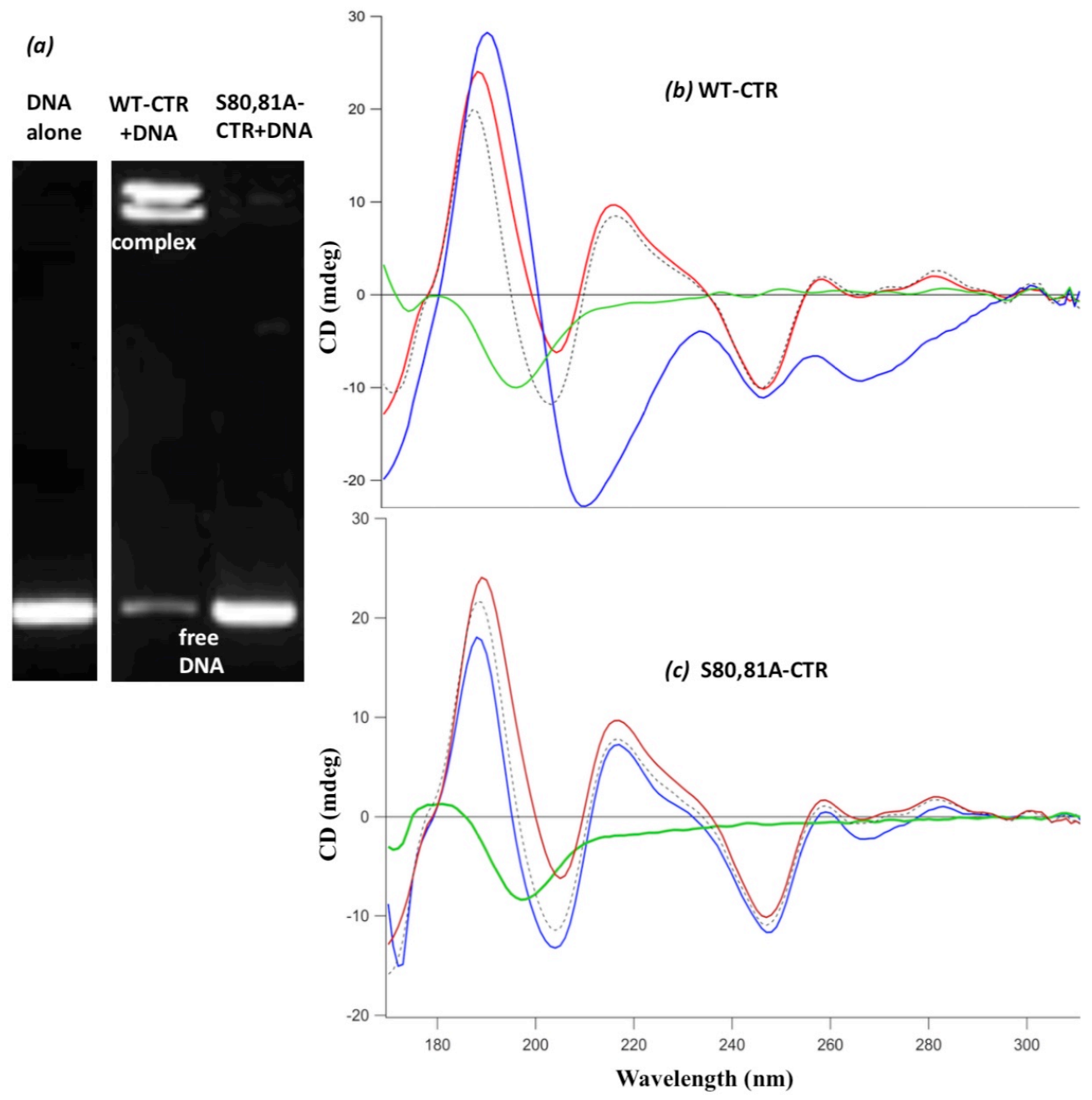

***Figure 1**: **(a)** EMSA analysis of Hfq-CTR wild type (WT-CTR) and mutant (S80,81A-CTR) in the presence of DNA. DNA alone; DNA in the presence of WT-CTR; DNA in the presence of S80,81A-CTR mutant. EMSA shows that the mutant does not bind to DNA, while WT-CTR does. Note that 2 complexes of different size are formed by the WT Hfq-CTR, probably due to different numbers of CTRs bound to DNA. **(b)** SRCD spectra of the WT-CTR/DNA complex (blue), DNA (red) and WT-CTR (green). The dotted spectrum represents the relevant combination of the spectra pertaining to the individual components. **(c)** As in panel (b) but for S80,81A-CTR.*

## DNA and CTR dispersions studied *via* wire µrheology

### *Generic behaviors: Viscous liquid versus viscoelastic solid*

Figure 2a illustrates the rotation of a 33 µm wire in a 7.3 mM $(dA{:}dT)_{59}$ DNA dispersion at angular frequency $\omega$ = 0.31 rad $s^{-1}$, with a counter-clock wise rotation in the direction indicated by yellow arrows. The different images of the chronophotograph are taken at time intervals of 2.4 s during a 210°-rotation of the object, showing that the wire rotates synchronously with the field. By increasing the frequency to 9.4 rad $s^{-1}$ (Figure 2b), the wire presents a transition to asynchronous rotation whereby it shows back and forth oscillations. After an initial increase in the orientation angle in the counter-clockwise direction (yellow arrows), the wire undergoes a back motion in the clockwise direction (red arrow) followed again by an increase in orientation angle.

We refer to the transition from the synchronous to the asynchronous regime by the critical frequency $\omega_C$. For the wire in Figure 2a and Figure 2b, $\omega_C$ is equal to 3.7 ± 0.5 rad $s^{-1}$. A 1.8 mM CTR-dispersion was studied in the same conditions and revealed a similar transition with increasing $\omega$. As reported in our work on model viscoelastic liquids and soft solids[45, 59], the previous behavior was found to be characteristic of purely viscous liquids.

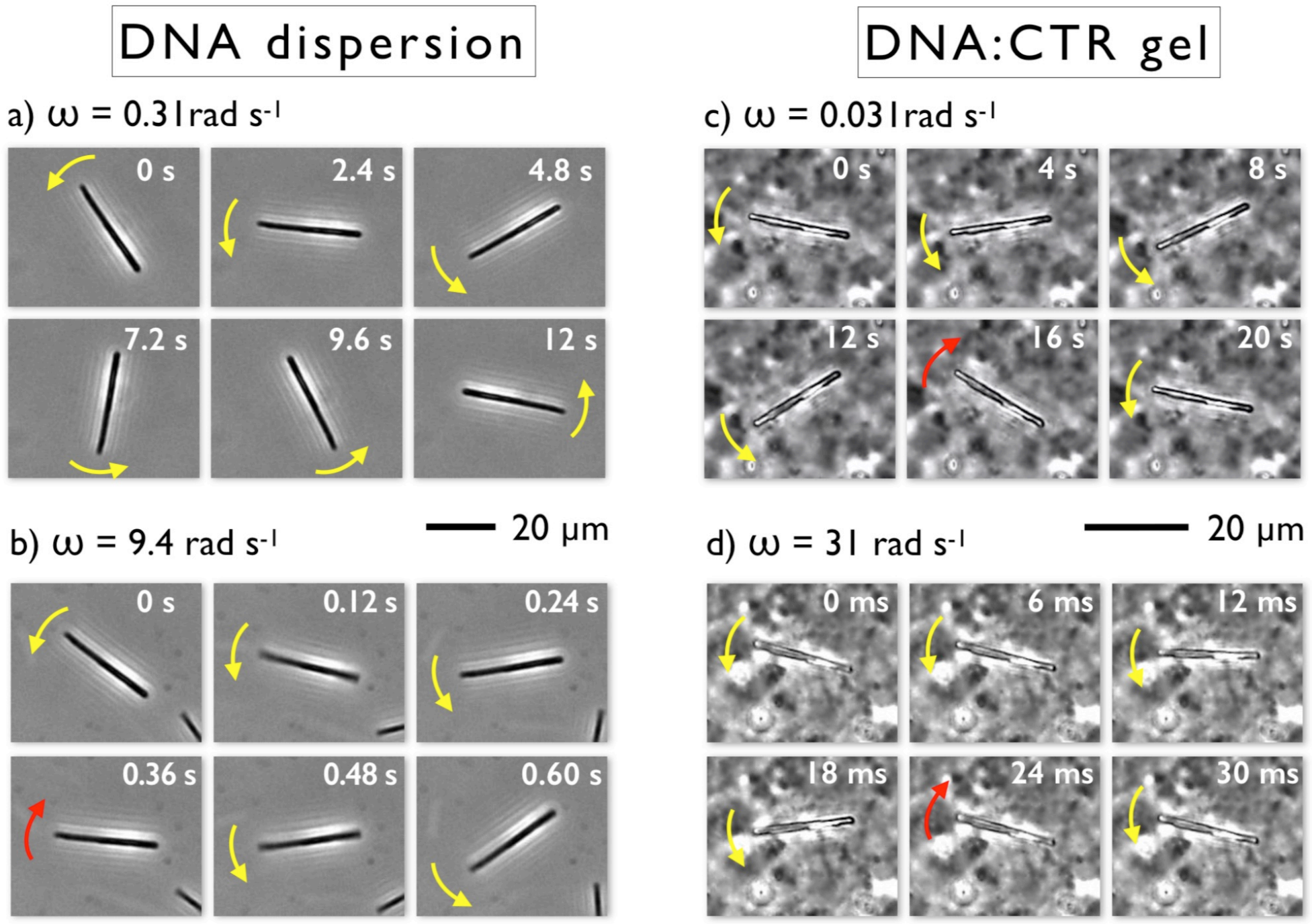

***Figure 2**: **(a)** Chronophotograph of a 33 µm (diameter 1.8 µm) wire undergoing a rotation observed in a 7.3 mM DNA dispersion under the application of a magnetic field $\mu_0 H$ = 10.3 mT and at room temperature. In this experiment, the applied actuation frequency $\omega$ is 0.31 rad $s^{-1}$. **(b)** As in Figure 2a at $\omega$ = 9.4 rad $s^{-1}$. The yellow (resp. red) arrows mark the rotation in the same (resp. opposite) direction of the magnetic field. **(c)** Chronophotograph of a 21 µm (diameter 1.4 µm) wire undergoing an oscillatory motion in a 7.3 mM:1.8 mM DNA-CTR complex at angular frequency $\omega$ = 0.031 rad $s^{-1}$. **(d)** As in Figure 2c at $\omega$ = 9.4 rad $s^{-1}$.*

For DNA-CTR complexes, similar measurements showed a notable change in wire oscillatory behavior. Subject to the same field conditions, the wires oscillated between fixed orientations, which were independent of angular frequency. Figure 2c and Figure 2d display chronophotographs of a 21 µm wire incorporated in a 4:1 complex and actuated at $\omega$ = 0.031 and 31 rad $s^{-1}$, respectively. One observes a heterogeneous background under phase contrast microscopy, which is associated with a non-uniform material distribution (as was shown by localized µrheology measured using bead technology, see below). This could come from heterogeneity in the formation of CTR assembled structures on DNA with sizes in the visible light range. It is also noted that the sizes of the heterogeneous regions (~ 1 µm) are smaller than the length of the wire. For this sample, the wire rotation is found to be asynchronous at all $\omega$'s, which implies that the wires do not follow the field completely for any applied actuation frequency. The observation on DNA-CTR is an indication that the material is a soft solid and characterized by a yield stress behavior. This outcome is also in line with results found in calcium ion crosslinked polysaccharide gels[47]. The videos associated with the wire rotations in Figure 2 are shown in Supporting Movies 1, 2, 3 and 4.

***Quantitative analysis***

For data analysis, movies of rotating wires were recorded, treated following the procedure described in the *Experimental Procedures* section and transformed into in-plane wire orientation angle $\theta(t)$ as a function of the time. The obtained $\theta(t)$ dependences were then analyzed in a set of two parameters: the average rotation velocity $\Omega(\omega) = \langle d\theta(t)/dt \rangle_t$ and the back-and-forth oscillation amplitude $\theta_B(\omega) = \langle \theta_B(t,\omega) \rangle_t$. Figure 3a shows the normalized velocity $\Omega(\omega)/\omega_C$ as a function of normalized frequency $\omega/\omega_C$ for both DNA and CRT dispersions at 7.3 and 1.8 mM. We find that $\Omega(\omega)/\omega_C$ increased with the frequency in the synchronous regime ($\omega \leq \omega_C$) until it passed through a sharp maximum at $\omega_C$ and then fell rapidly ($\omega > \omega_C$). One observes moreover that the wire response agrees well with the predictions of the Newton constitutive equation, given by $\Omega(X) = X$ for $X \leq 1$ and $\Omega(X) = X - \sqrt{X^2 - 1}$ for $X \geq 1$ (continuous line)[59].

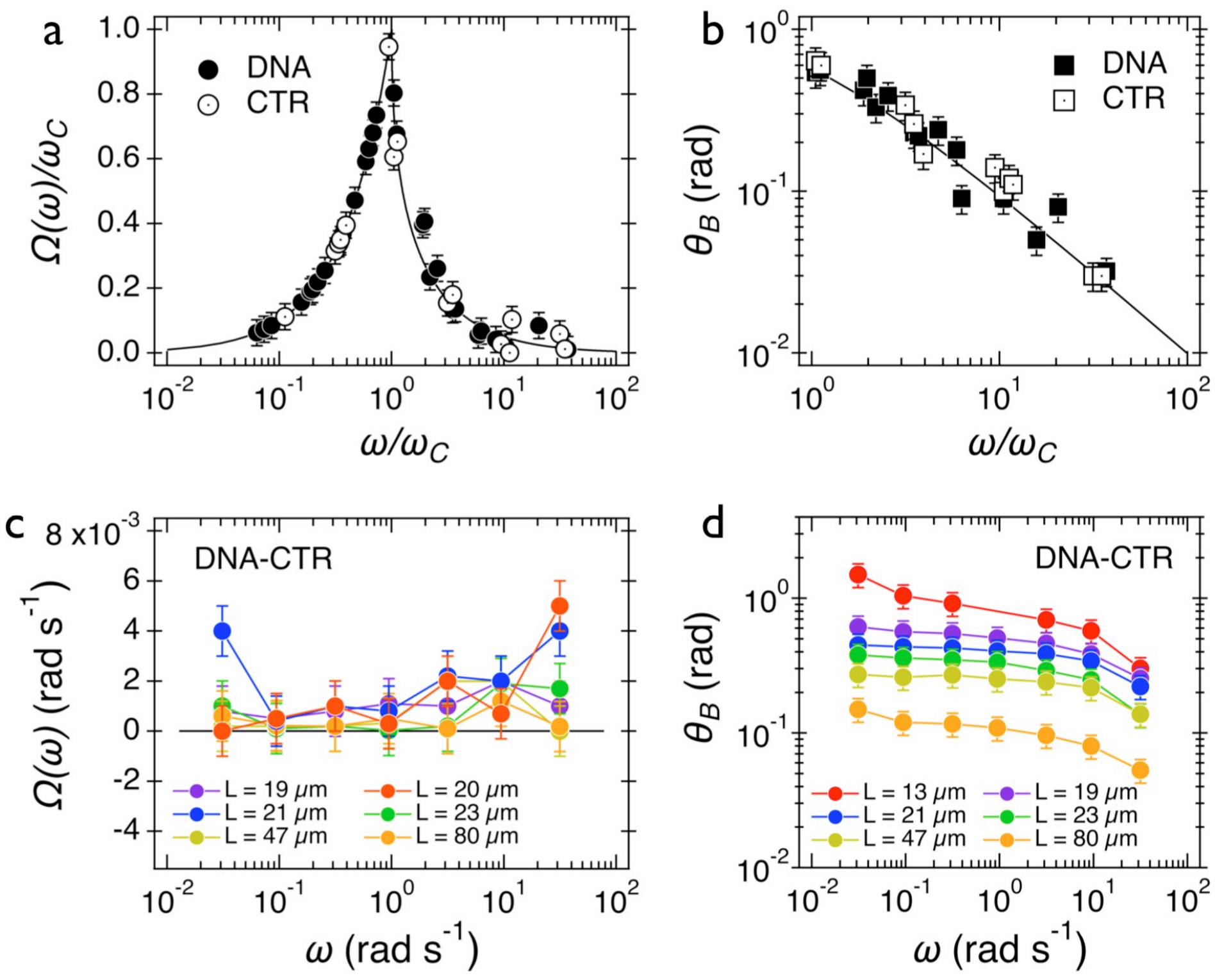

***Figure 3:*** ***(a, b)*** *Wire average rotation velocity $\Omega(\omega)$ and oscillation amplitude $\theta_B(\omega)$ measured in DNA and CTR dispersions. The continuous lines are calculated from the constitutive equations for Newton fluids.* ***(c, d)*** *Same as 2a and 2b for the 4:1 DNA-CTR complex. For the mixed dispersions, $\Omega(\omega)$ is close to zero and frequency independent. The oscillation amplitude displays a strong dependency with the wire length.*

Figure 3b shows the oscillation amplitude angle $\theta_B(\omega)$ as a function of the reduced frequency $\omega/\omega_C$ for the same 2 samples. A good superimposition of the data points is observed and is in agreement again with the Newton prediction, $\theta_B(\omega/\omega_C) \sim (\omega/\omega_C)^{-1}$ for $\omega \gg \omega_C$ (continuous line)[45, 60]. Note that in the $\theta_B(\omega/\omega_C)$-representation, the continuous line in the figure is obtained without an adjustable parameter[45]. In contrast, DNA and CTR behaved as Newton fluids at the concentrations studied here and the dispersions were characterized by a single rheological parameter, the static shear viscosity $\eta$, determined in the next section. Figure 3c displays the average angular velocity $\Omega(\omega)$ in 4:1 DNA-CTR dispersions for wires comprised between 19 and 80 µm and actuated in the range $\omega$ = 3×10$^{-2}$ – 30 rad s$^{-1}$. Data show that $\Omega(\omega)$ are frequency independent and close to zero. For this sample, the critical frequency $\omega_C$ cannot be found. Such behavior has been identified as resulting from soft solids, characterized by an infinite static viscosity and a yield stress. Figure 3d displays the oscillation amplitude $\theta_B(\omega)$ as a function of the frequency. $\theta_B(\omega)$ exhibits a plateau at low frequency, noted $\theta_{B,M}$ and decreases above 1 rad s$^{-1}$, showing again a very different behavior compared with Newton liquids (Figure 3b).

We next take advantage of the wire length dispersity and explore asymptotic behaviors, a technique that was shown to improve the accuracy in the determination of rheological parameters[52]. For (dA:dT)$_{59}$ and CTR dispersions, the viscosity can be identified from the asymptotic behavior of $\omega_C$ as a function of $L^*$, the reduced wire length. As shown previously[45], $\omega_C$ is related with the static viscosity $\eta$ *via* the expression: $\omega_C = 3\mu_0\Delta\chi H^2/8\eta {L^*}^2$, where $\mu_0$ the vacuum permeability, $H$ the magnetic field strength and $\Delta\chi$ = 0.054, the magnetic anisotropy of wires. Figure 4a displays the ratio $8\eta\omega_C/3\mu_0\Delta\chi H^2$ as a function of $L^*$ for the DNA, CTR, and a control consisting of a complex between DNA and an Hfq CTR mutant. This control peptide (mutant CTR) was used to show that the gelation is due specifically to the wild type Hfq-CTR properties and not to a nonspecific effect due to the presence of the protein only. This result is indicative that the Hfq-CTRS80,81A mutant is unable to bind DNA as was discussed above (Figure 1). On Figure 4a, the data points collapse on a single master curve displaying the $1/{L^*}^2$-dependence (continuous line) in agreement with the model predictions. The measured viscosities are 1.6 ± 0.3 mPa s for the DNA solution, 3.3 ± 0.8 mPa s for CTR and 2.4 ±0.3 mPa s for 4:1 DNA- CTRS80,81A. These values are slightly higher than the viscosity of the suspending solvent ($\eta_{PBS}$ = 0.90 mPa s at 25 °C) and in line with that of a dilute solution of macromolecules[61]. Given the low concentration in the CTR sample (1.8 mM), a viscosity of 3.3 mPa s may indicate possible aggregation of proteins.

From the constitutive equations developed for soft solids[47], the low frequency limit $\theta_{B,M}$ is linked to the equilibrium storage modulus $G_{eq}$ through the relation $\theta_{B,M} = \lim_{\omega\to 0}\theta_B(\omega) = 3\mu_0\Delta\chi H^2/4{L^*}^2 G_{eq}$. For a soft solid, $G_{eq}$ characterizes the quasi-static elastic response of the material. Figure 4b shows the ratio $4G_{eq}\theta_{B,M}/3\mu_0\Delta\chi H^2$ as a function of $L^*$ for two different samples of DNA-CTR gel. A good superimposition of the data points is observed and agrees with the predicted $1/{L^*}^2$ scaling. The equilibrium storage moduli are equal to 0.2 ± 0.03 Pa and 1.7 ± 0.3 Pa for these two experiments, the susceptibility anisotropy coefficients being $\Delta\chi$ = 2.3. The results suggest a certain variability in the modulus which could come from fluctuations in the elasticity of the environment or from sample aging effects. In conclusion and considering the fre-quency range studied, the 4:1 DNA-CTR complex behaves as a soft gel with a clear viscoelastic signature, whereas a complex made with the CTR mutant is a purely viscous liquid.

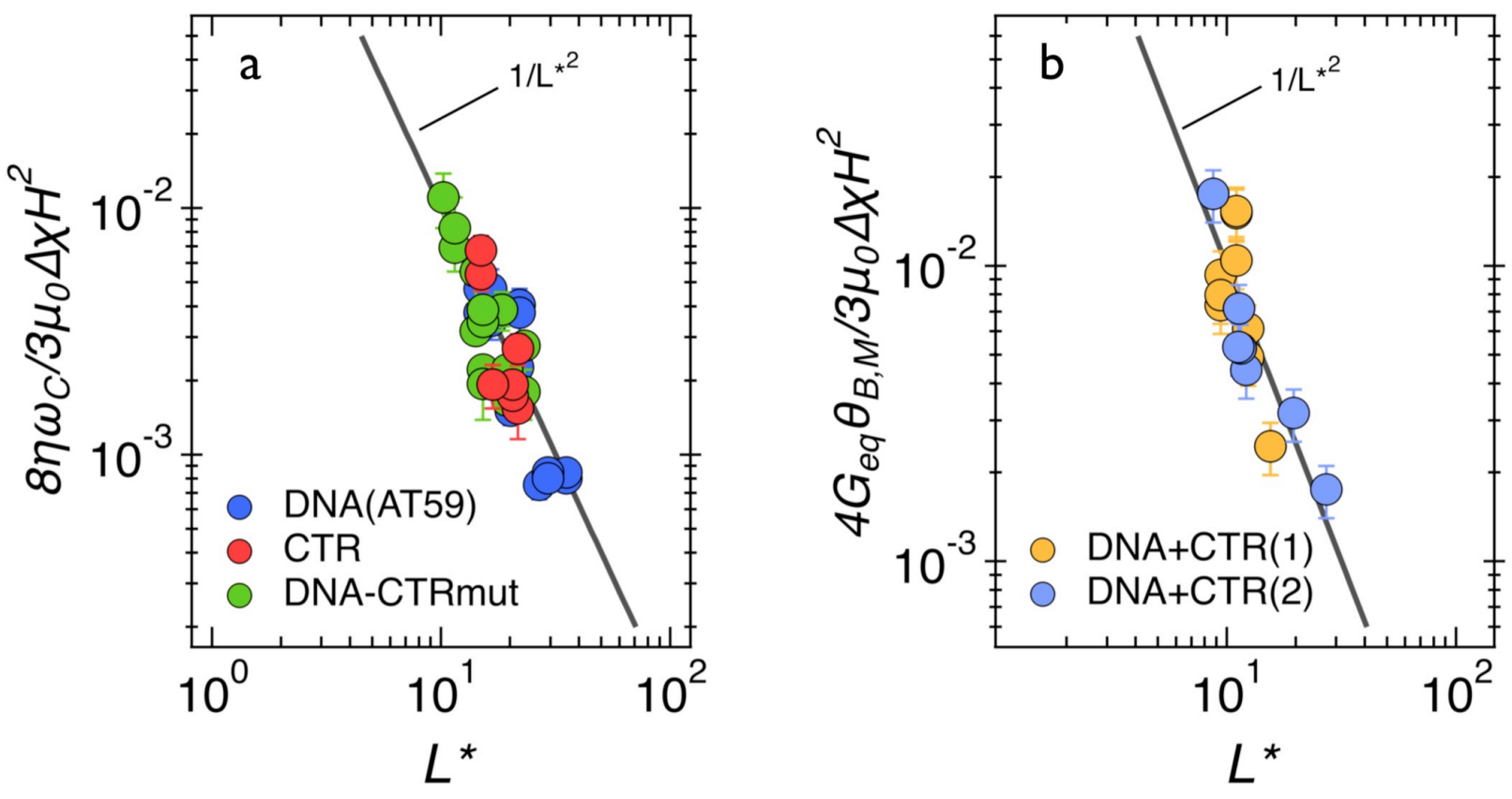

***Figure 4**: **(a)** Normalized critical frequency $8\eta\omega_C/3\mu_0\Delta\chi H^2$ as a function of the reduced wire length $L^*$ obtained for DNA, CTR and 4:1 DNA-Hfq-CTRS80,81A complex dispersions. The straight lines correspond to the $1/L^{*2}$ dependence predicted by the Newton constitutive equation. **(b)** Normalized oscillation amplitude $4G_{eq}\theta_{B,M}/3\mu_0\Delta\chi H^2$ as a function of $L^*$ for two different samples of 4:1 DNA-CTR complexes. The straight line indicates the $1/L^{*2}$-behavior expected from the soft solid model[47].*

## DNA and CTR dispersions studied via particle tracking

### *Passive bead microrheology of DNA and of DNA-CTR complexes*

The viscoelastic properties of the DNA-CTR dispersions were also measured with passive µrheology using video tracking of colloidal beads. This method allows the measurement of the viscoelastic response in localized regions of the sample, thereby characterizing the heterogeneity in CTR assembled structures on DNA. Furthermore, the dispersions are investigated close to equilibrium using low shear without causing variation in DNA density over several microns. Formation of a physical gel can easily be recognized from plateauing bead mean square displacement for long lag-times and its temperature reversible behavior.

From the trajectories of the colloidal beads, the probability distributions for displacements in the *x*- and *y*-directions were determined for a range of lag-times *t*. Gaussians were fitted to the probability distributions by optimizing the mean square displacement $<\Delta x^2(t)>$. The widths of the distributions obtained in the orthogonal *x* and *y* directions were averaged. Accordingly, $<\Delta x^2(t)>$ refers to the mean square displacement in one dimension. We then obtained the time-dependent creep compliance according to $J(t) = 3\pi a <\Delta x^2(t)>/k_B T$ with $k_B T$ being thermal energy and $a$ the radius of the bead. The results for $(dA{:}dT)_{59}$ and 4:1 DNA-CTR complex obtained at ambient temperature (296 K) are shown in Figure 5a and Figure 5b, respectively. Results for 1:1 and 2:1 DNA-CTR dispersions are qualitatively similar but differ in relaxation time and elasticity modulus (see below). The creep compliance for DNA shows viscous flow behavior according to $J(t) = t/\eta$ with static viscosity $\eta = 1.13 \pm 0.03$ mPa s. For the DNA-CTR disper-sions, the temporal behavior of the creep compliance is markedly different. Here, $J(t)$ first increases and, subsequently, plateaus at a constant value ($1/G'$, see below) with increasing lag-time. Furthermore, the creep compliance depends on the location of the monitored bead inside the sample (10-25 different locations were probed). We have verified that the variation in creep compliance is not related to a variation in bead diameter but to inhomogeneity in viscoelastic response.

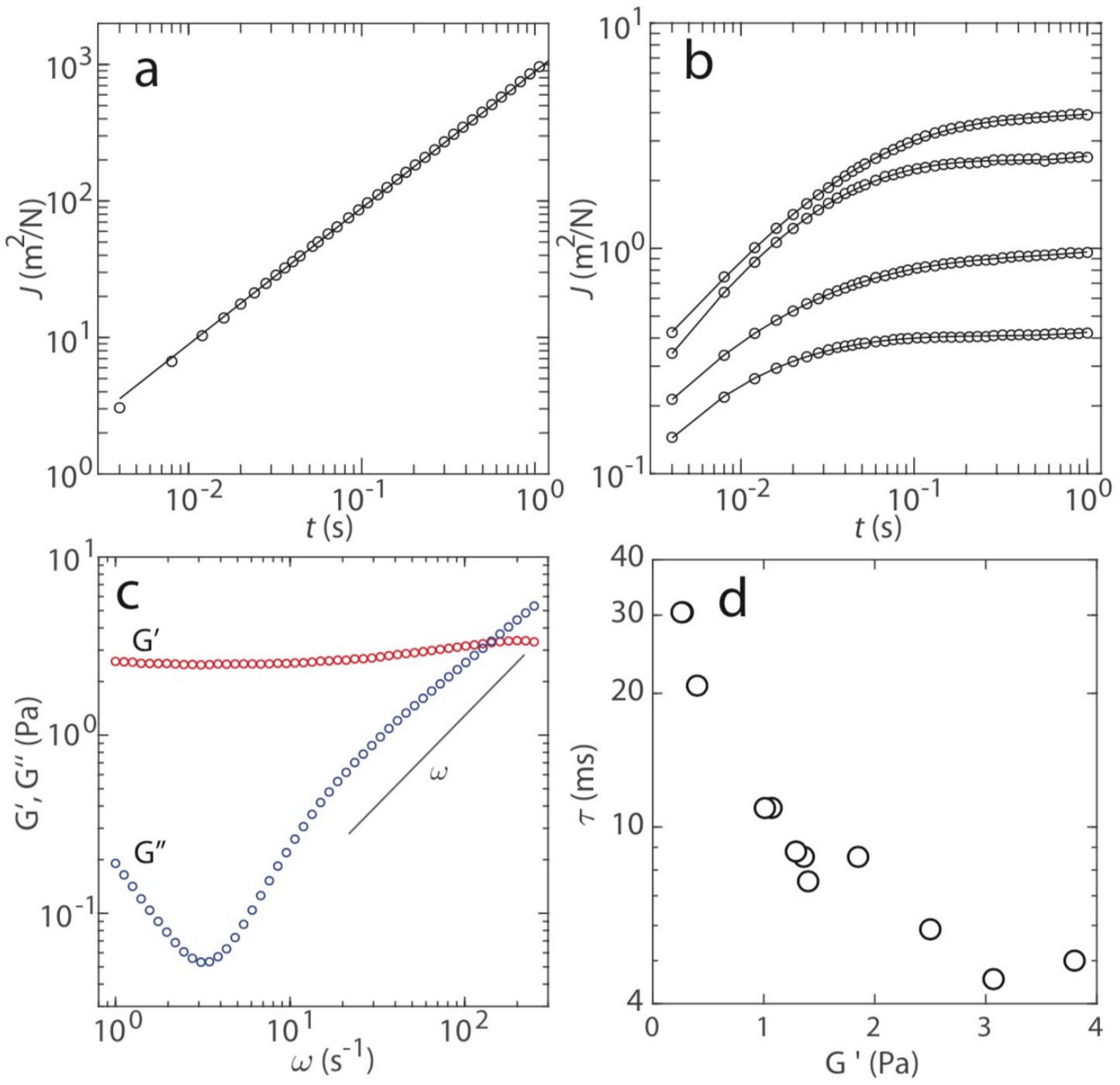

***Figure 5. (a)** Creep compliance J versus time t at 296 K of the model system without CTR. The DNA concentration is 1.8 mM. The solid line represents a linear fit. **(b)** As in panel **(a)**, but for 4:1 DNA-CTR dispersion at four randomly chosen bead locations inside the sample. **(c)** Elastic storage $G'$ and viscous loss $G''$ moduli versus frequency $\omega$ pertaining to the lowest creep compliance in panel **(b)**. The solid line denotes $\omega$ scaling. **(d)** Low frequency limit of $G'$ and cross-over relaxation time $\tau = {\omega_c}^{-1}$ pertaining to 12 different locations.*

The viscoelastic response of the DNA-CTR dispersions is more conveniently discussed in terms of the elastic storage and viscous loss moduli, $G'$ and $G''$, respectively. Following the previously described procedure, we have obtained $G'$ and $G''$, from the one-sided, complex Fourier transform of the creep compliance and the generalized Stokes-Einstein equation [62]. This approach requires that the fluid can be treated as an incompressible continuum, no-slip boundary conditions, and that the Stokes drag can be extended over all frequencies [63]. An example of the moduli is displayed in Figure 5c. If the bead is harmonically bound according to $< \Delta x^2(t) > = x_o^2\,[1 - \exp(-t/\tau)]$ with amplitude $x_o^2$ and relaxation time $\tau$, one obtains a constant $G' = k_B T/(3\pi a x_o^2) = 1/J(t \to \infty)$, $G'' = \omega\tau\, G'$, and a cross-over frequency $\omega_c = \tau^{-1}$. As can be seen in Figure 5c, $G'$ is almost constant and $G''$ approaches $\omega$ scaling for higher frequencies. Accordingly, the passive viscoelastic behavior of the DNA-CTR dispersions at the relevant time scale (< 1 s) is close to the one of an elastic solid. As is the case for the creep compliance, $G'$ and $G''$ vary across the sample. For different locations, we determined the low frequency limit of $G'$ and cross-over relaxation time $\tau = \omega_c{}^{-1}$. The results are shown in Figure 5d (12 locations). A variation over an order of magnitude is observed with a longer relaxation time for locations with lower values of $G'$. Similar spatial variation over different locations was observed for the other investigated DNA-CTR dispersions. At higher temperature, the elastic modulus decreases but the dispersions remain elastic solid-like. We have checked that the temperature induced change in viscoelasticity is reversible. The mean values and standard deviations of $G'$ and $\tau$ pertaining to the various investigated DNA-CTR dispersions are collected in Table II. Note that the data were obtained by averaging results for 10 - 25 different locations across each sample, so that the standard deviations refer to inhomogeneity in viscoelastic response rather than experimental error.

| Sample | Composition | T (K) | $G'$ (Pa) | $\tau$ (ms) |
|---|---|---|---|---|
| 1:1 DNA-CTR | 1.8 mM DNA 1.8 mM CTR | 296 | 1.4 (0.8) | 11 (5) |
| | | 303 | 1.1 (0.8) | 13 (10) |
| | | 313 | 0.9 (0.7) | 13 (9) |
| | | 323 | 0.6 (0.5) | 26(20) |
| | | 333 | 0.03 (0.02) | 104 (60) |
| 2:1 DNA-CTR | 7.8 mM DNA 3.9 mM CTR | 296 | 2.6 (1.0) | 7 (2) |
| | | 303 | 1.9 (0.8) | 6 (1) |
| | | 313 | 1.3 (0.7) | 10 (4) |
| | | 323 | 0.7 (0.4) | 11 (5) |
| | | 333 | 0.2 (0.1) | 23 (14) |
| 4:1 DNA-CTR | 7.3 mM DNA 1.8 mM CTR | 296 | 1.5 (1.1) | 13 (9) |

***Table II***: *Mean values of the low frequency elasticity modulus $G'$, relaxation time $\tau$ and their standard deviations (bracketed) at 10-25 bead locations across the DNA-CTR dispersions.*

The elasticity moduli for 1:1 and 4:1 DNA-CTR complexes are similar and comparable to those which have been measured with wire rheology. An increase in concentration of DNA but with constant concentration of CTR has little, if any effect on the rheological properties. On the other hand, doubling the CTR concentration at approximately the same DNA concentration (2:1 *vs* 4:1 complex, Table II) results in a gel with a higher elasticity modulus and a shorter relaxation time. These observations indicate that the compliance of the system is determined by the density of crosslinks formed by complexation of CTR on DNA. An increase in temperature results in a decrease in crosslink density (cross-linking complexes are reversibly dissolved), larger but less dynamic units, and, hence, a decrease in elasticity modulus. At the sub-second time scale of the passive rheology experiment, the DNA-CTR dispersions behave close to an elastic solid. The derived relaxation times refer, hence, to thermally activated elastic deformations of the cross-linked network of DNA and CTR. For longer times the dispersions might flow with relaxation times exceeding tens of seconds. The latter dynamics likely involves crosslink rearrangements.

Unfortunately, such slow dynamics are outside the window of observation of the particle tracking experiment due to long time experimental artifacts such as drift of the microscope stage.

Finally, samples with relatively high concentrations of DNA and Hfq-CTR (6.5 mM each) exhibited separation into two coexisting phases (in agreement with non-uniform material distribution observed previously under phase contrast microscopy, see Figure 2). One phase was a Newtonian fluid, whereas the other phase behaved as a gel. The viscoelastic response of the gel could not be measured because the beads were almost immobilized. This phase separation may help to compact the genomic DNA *in vivo*, as previously suggested by *in silico* analyses [7].

## CONCLUSION

The key finding of this work is that a putative DNA binding domain of a Nucleoid Associated Protein assembles DNA into a gel. The interaction of DNA with the amyloid domain has been revealed by SRCD spectroscopy. The properties of the resulting gel may either be a true elastic soft solid with a yield stress value of a viscoelastic liquid with a finite viscosity. Our experiments rather suggest that it is a soft solid, the creep compliance as measured by bead tracking plateaus at a constant value for lag-times of around 1 s and the lowest frequency that could be reached with the wire rheology is 0.01 rad/s. If the network has a relaxation time associated with yield to flow, it is very long (> 100 s) for the range of shear stresses applied in this work. Both wire rheology with the employed range in actuation frequency and bead tracking methodology probe the linear response of the material. As shown previously using solutions of wormlike micelles and gels made from cross-linked polysaccharides, the static viscosity determined from the wire-based rheology and that obtained from standard cone-and-plate measurements are in good agreement, indicating that actuated wires are probing the linear response of the material [45]. The passive thermal motion of the embedded colloidal beads does not induce a variation in DNA density over distances exceeding their temporal displacements, that is several microns. The observed inhomogeneity in elasticity modulus originates from variation in cross-link density, which is a characteristic property of the solution and not caused by the probing beads. In the case of wire rheology, the DNA-CTR gel deforms but does not flow or break. This is due to the fact that the stress applied by the wire is below the yield stress value of the gel. The elastic moduli obtained with both techniques agree and are around 1 Pa. Bead rheology shows however a variation over an order of magnitude across sample, highlighting inhomogeneity in density of crosslinks formed by dynamic complexation of CTR on DNA. An increase in temperature results in a reversible decrease in crosslink density and softening of the gel.

As for *in vivo* consequences, our experiments suggest that a NAP not only determines the spatial organization of the genome but also properties of the flow. The latter aspect is often overlooked but should have important consequences for understanding molecular motion of DNA and proteins in the context of the machinery of life[6]. The assembly of such DNA-protein complexes is usually rather slow, as the kinetics is limited by the diffusion of the constituents (low concentrations). Therefore, this micromechanical constraint generated by Hfq, even if it should be very local, could serve various biological processes. DNA's viscoelasticity might have impli-cations for cellular processes, in particular those that involve transport and flow of the genome. In particular, cell division requires separation of the replicated genetic material into the daughter cells. Other examples are protein-mediated looping of chromosomal sections, chromosomal organization in topological domains[64], and transcriptional bursting [65-68]. We surmise that DNA's viscoelasticity also provides a new role for NAPs in protein target search, because a protein diffusing through a network of DNA is affected by viscous forces (hydrodynamic interaction) and requires DNA segments moving out of the way. Another in vivo consequence of gelation might be swelling of the DNA network by an imbalance in osmotic pressure, resulting in stress exerted on duplex and concomitant effects on protein binding.

## AUTHOR INFORMATION

* Corresponding author: Véronique Arluison; LLB UMR 12 CEA/CNRS, CEA Saclay, 91191 Gif-sur-Yvette Cedex, France; Tel 33 (0)1 69 08 32 82; Email: veronique.arluison@u-paris.fr

[†] Present Address: Omar El Hamoui, Institut des Sciences Analytiques et de Physico-chimie pour l'Environnement et les Matériaux, UMR 5254 CNRS/Université de Pau et des Pays de l'Adour, 2 Avenue du Président Pierre Angot, 64000 Pau, France

## AUTHOR CONTRIBUTIONS

[#] OEH and IY contributed equally to the work and are equal first authors. JvdM, JFB, FW and VA conceived the original idea and designed the experiments. OEH, IY, VA and FW performed experiments. JvdM, JFB, FW, MR and VA wrote the manuscript. All authors analysed results and commented on the manuscript.

## FUNDING SOURCES

This work was supported by CNRS, synchrotron SOLEIL and CEA. ANR (Agence Nationale de la Recherche) and CGI (Commissariat à l'Investissement d'Avenir) are gratefully acknowledged for their financial support of this work through Labex SEAM (Science and Engineering for Advanced Materials and devices) ANR 11 LABX 086, ANR 11 IDEX 05 02. We acknowledge the ImagoSeine facility (Jacques Monod Institute, Paris, France), and the France BioImaging infrastructure supported by the French National Research Agency (ANR-10-INSB-04, « Investments for the future »). This research was supported in part by the Agence Nationale de la Recherche under the contract ANR-13-BS08-0015 (PANORAMA), ANR-12-CHEX-0011 (PULMONANO), ANR-15-CE18-0024-01 (ICONS), ANR-17-CE09-0017 (AlveolusMimics) and by Solvay. MR acknowledges receipt of financial support from Université de Paris under Bourse Qualite de Recherche BQR/UFR de physique (number PADYF16RER, 2019). This research was supported in part by Singapore Ministry of Education Academic Research Fund Tier 1.

## ACKNOWLEDGMENT

We thank R.R. Sinden (SDSMT, USA) and W. Grange (U. of Paris) for fruitful discussions and comments on the manuscript. We are grateful to D. Partouche (Synchrotron SOLEIL and LLB) for his contribution to µrheology measurements at an early stage of this work and to F. Turbant (LLB) for his help in mutant characterization. SRCD measurements on DISCO beamline at the SOLEIL Synchrotron were performed under proposal 20171061.

**ABBREVIATIONS**
bp: base pair; CTR/NTR: Hfq C/N-terminal region; EMSA: Electrophoretic Mobility Assay; NAP: Nucleoid Associated Protein; µrheology: microrheology ; SRCD: Synchrotron Radiation Circular Dichroism; WT: wild type